\documentclass[journal]{IEEEtran}
\usepackage{graphicx}
\usepackage{caption}
\usepackage{tabularx,booktabs}

\usepackage{color}
\usepackage{multicol}
\usepackage{amsmath,amssymb}
\usepackage{amsfonts}
\usepackage{textcomp}
\usepackage{psfrag}
\usepackage{multimedia}
\usepackage{fancybox}

\newcommand{\E}[2]{\mathbb{E}_{#1}\left[#2\right]}
\newcommand{\Prob}[2]{\mathbb{P}_{#1}\left[#2\right]}

\newtheorem{theorem}{Theorem}

\newtheorem{Corollary}{Corollary}
\newtheorem{Lemma}{Lemma}

\newcommand {\matr}[2]{\left[\begin{array}{#1}#2\end{array}\right]}
\newcommand {\cmatr}[2]{\left\{\begin{array}{#1}#2\end{array}\right.}
\newcommand{\tightMatr}[2]{\begin{bmatrix}#2\end{bmatrix}}

\newcounter{lastnote}

\begin{document} 

\title{Data-driven Economic NMPC using Reinforcement Learning} 

\author
{S\'ebastien Gros, Mario Zanon
	\thanks{S.Gros is with the Department of Signal \& System, Chalmers University of Technology, H\"{o}rsalsv\"{a}gen 9a, G\"{o}teborg, Sweden.}
	\thanks{Mario Zanon is with the IMT School for Advanced Studies Lucca, Lucca
		55100, Italy.}
}



%
%

\IEEEtitleabstractindextext{
	\begin{abstract}
Reinforcement Learning (RL) is a powerful tool to perform data-driven optimal control without relying on a model of the system. However, RL struggles to provide hard guarantees on the behavior of the resulting control scheme.
In contrast, Nonlinear Model Predictive Control (NMPC) and Economic NMPC (ENMPC) are standard tools for the closed-loop optimal control of complex systems with constraints and limitations, and benefit from a rich theory to assess their closed-loop behavior. Unfortunately, the performance of (E)NMPC hinges on the quality of the model underlying the control scheme. In this paper, we show that an (E)NMPC scheme can be tuned to deliver the optimal policy of the real system even when using a wrong model. This result also holds for real systems having stochastic dynamics. This entails that ENMPC can be used as a new type of function approximator within RL. Furthermore, we investigate our results in the context of ENMPC and formally connect them to the concept of dissipativity, which is central for the ENMPC stability. Finally, we detail how these results can be used to deploy classic RL tools for tuning (E)NMPC schemes. We apply these tools on both a classical linear MPC setting and a standard nonlinear example from the ENMPC literature. 
	\end{abstract}
	
}

\maketitle

\IEEEdisplaynontitleabstractindextext

\IEEEpeerreviewmaketitle

\section{Introduction}

Reinforcement Learning (RL) is a powerful tool for tackling Markov Decision Processes (MDP) without depending on a model of the probability distributions underlying the state transitions. Indeed, most RL methods rely purely on observed state transitions, and realizations of the stage cost in order to increase the performance of the control policy. RL has drawn an increasing attention thanks to its striking accomplishments ranging from computers beating Chess and Go masters \cite{44806}, to robot learning to walk or fly without supervision \cite{Wang:2012:MLA:2719652.2719704,Abbeel07anapplication}. 

Most RL methods are based on learning the optimal control policy for the real system either directly, or indirectly. Indirect methods rely on learning an approximation of the optimal action-value function underlying the MDP, typically using variants of Temporal-Difference learning \cite{Kober2013}. Since the action-value function is in general unknown a priori, a generic function approximator is typically used to approximate it. A common choice in the RL community is to use a Deep Neural Network (DNN).  

Direct RL methods seek to learn the optimal policy directly. Most direct RL methods are based on the stochastic or deterministic policy gradient methods, see e.g. \cite{Sutton:1999:PGM:3009657.3009806, Silver2014}. Both rely on carrying an approximation of the action-value function, or at least of the value function underlying the policy. Similarly to indirect methods, also direct RL methods typically use DNNs to approximate the optimal policy and the associated (action-) value function. 

Unfortunately, the closed-loop behavior of a system subject to an approximate optimal policy supported by a DNN or a generic function approximation can be difficult to formally analyze. It can therefore be difficult to generate certificates of the behavior of a system controlled by a generic RL algorithm. This issue is especially salient when dealing with safety-critical systems. The development of safe RL methods, which aims at tackling this issue, is an open field or research \cite{Garcia2015}.


Nonlinear Model Predictive Control (NMPC) is a formal control method based on solving at every time instant an optimal control problem to generate the optimal control policy. The optimal control problem seeks to minimize a sum of stage costs over a prediction horizon, subject to state trajectories provided by a model of the real system, the current observed state of the real system, and possibly state and input constraints to be respected. The optimal control problem then delivers an entire input and state sequence spanning the prediction horizon. However, only the first input is applied to the real system. At the next time instant, the entire optimal control problem is solved again using a new estimation of the state of the system.

If the system model underlying the NMPC scheme is perfect, and with the addition of an adequate terminal cost, the NMPC scheme delivers the optimal control policy. Classic NMPC is based on a stage cost lower-bounded by $\mathcal{K}_\infty$ functions, while Economic NMPC (ENMPC) accepts a generic stage cost \cite{Rawlings2009,Diehl2011,Amrit2011a}. A rich and mature theory exists in the literature to analyze the properties of classic NMPC when operating in closed-loop with a real system, establishing desirable key properties such as recursive feasibility and stability \cite{Mayne2000a,Rawlings2009b,Grune2011}. ENMPC has recently attracted the attention of the research community, and a stability theory has been developed fairly recently \cite{Rawlings2009,Diehl2011,Amrit2011a,Angeli2012a,Mueller2015}, but is arguably still under development. Since it can tackle the system constraints directly and since it benefits from a rich theory analyzing its behavior, (E)NMPC is arguably an ideal candidate for safety-critical applications.

Unfortunately, the performance of (E)NMPC schemes relies on having a good model of the system to be controlled. A data-driven adaption of the NMPC model to better fit the real system is a fairly obvious approach to tackle this issue. However, since the model does not necessarily match the real system, fitting the model to the data does not necessarily result in the (E)NMPC scheme delivering the optimal policy, and can even be counterproductive. Some attempts have been recently proposed to tackle this problem such as e.g. in \cite{Bla}.

The problem of optimizing a system based on a model having the wrong structure is well known in the field of Real-Time Optimization \cite{Forbes1994,Forbes1996,Agarwal1997,Gao2005a,Marchetti2009}, and has been addressed via the Modifier Approach \cite{Roberts1979,Roberts1995,Tatjewski2002,Gao2005b,Marchetti2009}, whereby the cost function of the optimization problem is adapted rather than the model. 

In this paper, we propose to use (E)NMPC schemes instead of DNNs to support the parametrization to approximate the \mbox{(action-)}value functions and the policy. Similarly to the idea originated in RTO, we show that the NMPC scheme can deliver the optimal control policy even if the underlying model is incorrect, by adapting the stage cost, terminal cost and constraints only. This observation is applicable to any NMPC scheme such as e.g.  classic NMPC, ENMPC, robust and stochastic ENMPC. Furthermore, we establish strong connections between this cost adaptation and the concept of strict dissipativity, which is fundamental to the stability theory of ENMPC. 

One practical outcome of the theory proposed in this paper is that all RL techniques can be directly used to tune the NMPC scheme to increase its performance on the real system. Because the theory proposed is very generic, this observation holds e.g. for a stochastic system being controlled by an (E)NMPC scheme based on a deterministic model or a robust NMPC scheme using a scenario tree. Another practical outcome of the theory proposed in this paper is that if a stage cost attached to a given system yields a stabilizing optimal control policy, then an ENMPC scheme with a positive stage cost can in principle be tuned to deliver the optimal policy. A last practical outcome of the proposed theory is that using (E)NMPC as a parametrization for RL instead of DNN allows one to use the rich theory underlying (E)NMPC schemes in the context of RL, and e.g. deliver certificates on the behavior of the policy resulting from the learning process.

The combination of learning and control techniques has been proposed in e.g.~\cite{Koller2018,Aswani2013,Ostafew2016,Berkenkamp2017}. To the best of our knowledge, however, our paper is the first work (a) proposing to use NMPC as a function approximator in RL and (b) investigating the connection between RL and economic MPC.

The paper is structured as follows. Section \ref{sec:ZeSection} establishes the fundamental result of the paper, showing that an (E)NMPC scheme based on the wrong model can, under some conditions, nonetheless deliver the optimal control policy. Section \ref{sec:NMPCusingRL} details how these results can be used in practice. Section \ref{eq:RLENMPC} details how some classic RL techniques can be deployed to adjust the (E)NMPC parameters. Section \ref{sec:ENMPC} further develops the theory and details its connection to the fundamental concept of strict dissipativity underlying the stability theory of ENMPC, and details its consequences for using ENMPC as a parametrization for RL. Section \ref{sec:LQR} deploys the theory on the case of LQR with Gaussian noise, for which all the objects discussed in the theory can be built explicitly, and their practical meaning assessed. Section \ref{sec:Examples} proposes some illustrative examples.

\section{Optimal policy based on an inexact model} \label{sec:ZeSection}
We will consider that the real system we want to control is described by a discrete Markov-Decision Process (MPD) having the (possibly) stochastic state transition dynamics 
\begin{align}
\Prob{}{s_+\,|\, s,a}, \label{eq:TrueDynamics}
\end{align}
where $s,a$ is the current state-input pair and $s_+$ is the subsequent one. We will label $L(s,a)$ the stage cost associated to the MDP, possibly infinite for some state-input pairs $s,a$, which we will assume can take the form:
\begin{align}
\label{eq:OriginalStageCost}
L\left(s,a\right) &= l \left(s,a\right) +\mathcal{I}_\infty\left(h\left(s,a\right)\right)+ \mathcal{I}_\infty\left(g\left(a\right)\right)
\end{align}
where we use the indicator function:
\begin{align}
\mathcal{I}_\infty(x) = \cmatr{cc}{\infty &\text{if }x_i >0\text{ for some }i \\ 0&\text{otherwise}}.
\end{align}
In \eqref{eq:OriginalStageCost}, function $l$ captures the cost given to different state-input pairs, while the constraints
\begin{align}
\label{eq:FiniteL}
g(a) \leq 0,\quad h(s,a)\leq 0
\end{align}
capture undesirable state and inputs, and infinite values are given to $L$ when \eqref{eq:FiniteL} is violated. Note that we have separated pure input constraints and mixed constraints for reasons that will be made cleared later on.

With the addition of a discount factor $0<\gamma \leq 1$, \eqref{eq:TrueDynamics} and \eqref{eq:OriginalStageCost} yield the optimal policy $\pi_\star\left(s\right)$. Note that the notation \eqref{eq:TrueDynamics} is standard in the literature on MDPs, while the control literature typically uses the notation $s_+=f(s,a,\zeta)$, where $\zeta$ is a stochastic variable and $f$ a possibly nonlinear function. 
The  action-value function $Q_\star$ and value function $V_\star$ associated with the MDP 
are defined by the Bellman equations~\cite{Bertsekas2005}:
\begin{subequations}
\label{eq:Bellman}
\begin{align}
Q_\star\left(s,a\right) &= L\left(s,a\right) + \gamma \mathbb{E}\left[V_\star(s_+)\,|\, s,a\right], \label{eq:Bellman1}\\
V_\star\left(s\right) &= Q_\star\left(s,\pi_\star\left(s\right)\right) = \min_{a}\, Q_\star\left(s,a\right). \label{eq:Bellman2}
\end{align}
\end{subequations}
Throughout the paper we will assume that the MDP underlying the real system, the associated stage cost $L$ and the discount factor $\gamma$ yield a well-posed problem, i.e. the value functions defined by \eqref{eq:Bellman} are well-posed, and finite over some sets.

We then consider a model of the real system having the state transition dynamics
\begin{align}
\Prob{}{\hat s_+\,|\, s,a}, \label{eq:ModelDynamics}
\end{align}
which typically do not match \eqref{eq:TrueDynamics} perfectly. Note that \eqref{eq:ModelDynamics} trivially includes deterministic models as a special case.
Consider a stage cost defined as
\begin{align}
\hat L\left(s,a\right) =   \cmatr{cc}{Q_\star\left(s,a\right) - \gamma \mathcal{V}^+\left( s,a\right) &\text{if} \,\, \left|\,\mathcal{V}^+\left(s,a\right)\,\right| < \infty \\ \infty &\text{otherwise}} ,\label{eq:hatL_def}
\end{align} 
where $\mathcal{V}^+\left(s,a\right) = \mathbb{E}\left[V_\star\left(\hat s_+\right)\,|\, s,a\right]$ and where the expectation is taken over the distribution \eqref{eq:ModelDynamics}. 
It is useful here to specify that the conditional formulation of the stage cost $\hat L$ proposed in \eqref{eq:hatL_def} is a technicality dedicated to having a well-defined $\hat L$ even when both the action value function and value function take infinite values.

We establish next the central theorem of this paper, stating that under some conditions the optimal policy $\pi_\star$ that minimizes the stage cost $L$ for the true dynamics \eqref{eq:TrueDynamics} is also generated by using model \eqref{eq:ModelDynamics} combined with the stage cost $\hat L$. Hence it is possible to generate the optimal policy based on a wrong model by modifying the stage cost. It may be useful to specify here that our approach further in the paper will be to bypass the possibly difficult evaluation of \eqref{eq:hatL_def}, and replace it by learning $\hat L$ directly from the data, see Section \ref{eq:RLENMPC}.


\begin{theorem} \label{thm:ZeTheorem}
	Consider the optimal value function 
\begin{align}
\hat V_N(s) =\min_{\pi}\E{}{\gamma^N V_\star(\hat s^{ \pi}_N) + \sum_{k=0}^{N-1}\, \gamma^k \hat L(\hat s^{ \pi}_k, \pi(\hat s^{ \pi}_k))} \label{eq:VhatN:Definition}
\end{align}	
associated to the stage cost \eqref{eq:hatL_def}, the state transition model \eqref{eq:ModelDynamics}, and the terminal cost $V_\star(s)$ over an optimization horizon $N$. Here we define $\hat s^{\pi}_{0,\ldots,N}$ as the (possibly stochastic) trajectories of the state transition model $\Prob{}{\hat s_+\,|\, s,a}$ under a policy $\pi$, starting from $\hat s^{ \pi}_0 = s$. We will label $\hat\pi$ the optimal policy associated to $\hat V_N(s)$ and $\hat Q_N(s,a)$ the associated action-value functions. Consider the set $\mathcal{S}$ such that
		\begin{align}
		\left|\,\mathbb{E}\left[V_\star\left(\hat s^{\pi_\star}_{k}\right)\right]\,\right| < \infty, \qquad \forall \, s\in\mathcal{S}, \qquad \forall \, k. \label{eq:ZeAssumption} 
		\end{align}
Then the following identities hold on $\mathcal{S}$:
	\begin{itemize}
		\item[(i)] 
		$\hat V_N(s) = V_\star(s)$ 
		\item[(ii)] $
		\hat\pi\left(s\right) = \pi_\star\left(s\right)$
		\item[(iii)] $
		\hat Q_N\left(s,a\right) = Q_\star\left(s,a\right)$ for the inputs $a$ such that $\left |\,\mathbb{E}\left[V_\star\left(\hat s_+\right)\,|\, s,a\right] \,\right | < \infty$.
	\end{itemize} 
\end{theorem}
\begin{IEEEproof} Let us consider the $N$-step value function $\hat V_N^\pi$ associated to the stage cost $\hat L$, the state transition model $\Prob{}{\hat s_+\,|\, s,a}$ and a policy $\pi$, defined as
	\begin{align}
		\hat V^\pi_N(s) &=\,\E{}{\gamma^NV_\star(\hat s_N^\pi) + \sum_{k=0}^{N-1} \gamma^{k} \hat L(\hat s_k^\pi,\pi(\hat s_k^\pi)) }. \label{eq:Vhat:Def}
			\end{align}
		Assumption \eqref{eq:ZeAssumption} ensures that, at least for $\pi=\pi_\star$, all the terms in the sum in \eqref{eq:Vhat:Def} have a finite expected value for $s\in\mathcal{S}$, such that $\hat V^\pi_N(s)$ is well defined and finite over $\mathcal{S}$ for some policies $\pi$. Using a telescopic sum, we can rewrite \eqref{eq:Vhat:Def} as:
			\begin{align}
	\hat V^\pi_N(s) 	&=\,Q_\star(s,\pi(s)) + \mathbb{E}\Bigg [ \sum_{k=1}^{N-1} \gamma^k A_\star(\hat s_k^\pi,\pi(\hat s_k^\pi))  \Bigg ], \label{eq:Vhat:Telescopic}
\end{align}
	where the advantage function $A_\star$ is defined as:
	\begin{align}
A_\star(s,a) =\cmatr{cc}{Q_\star(s,a) - V_\star(s)&\text{if }|Q_\star(s,a)|<\infty \\ 
\infty&\text{otherwise}}.
\end{align}
Using the Bellman equalities:
\begin{subequations}
\label{eq:Bellman:Identities}
\begin{align}
\min_{\pi}\,Q_\star(s,\pi(s)) &= Q_\star(s,\pi_\star(s)) = V_\star(s),\label{eq:Bellman:Identities:1} \\
\min_{\pi}\,A_\star(s,\pi(s)) &= A_\star(s,\pi_\star(s)) = 0,
\end{align}	
\end{subequations}
we observe that all terms in \eqref{eq:Vhat:Telescopic} are minimized by the policy $\pi_\star$, such that the following equalities hold on $\mathcal{S}$:
\begin{align}
\label{eq:TrickyEqualitiesInProof}
\hat V_N (s) = \min_\pi \hat V^\pi_N(s) = \hat V^{\pi_\star}_N(s) =V_\star(s),  
\end{align}
where the first equality holds by definition, the second holds because $\pi_\star$ is the minimizer of $\hat V^\pi_N$, and the last equality holds from \eqref{eq:Bellman:Identities}.
It follows that $\hat V_N (s) = V_\star(s)$ holds on $\mathcal{S}$ for any $N$, hence the identities (i) and (ii) hold. We furthermore observe that on $\mathcal{S}$ and for any input $a$ such that $|\mathbb{E}\left[V_\star\left(\hat s_+\right)\,|\, s,a\right]| < \infty$, the following equalities hold: 
\begin{align}
\hat Q_N\left(s,a\right) &=  \hat L\left(s,a\right) + \gamma\E{}{\hat V_{N-1}(\hat s_+)\,|\,s,a}\nonumber \\
&= Q_\star\left(s,a\right) + \gamma\E{}{\hat V_{N-1}(\hat s_+)-V_\star\left(\hat s_+\right)\,|\,s,a}\nonumber \\
&= Q_\star\left(s,a\right),
\end{align}
which yields statement (iii).
\end{IEEEproof}

	Note that if the transition model is exact, i.e. $\Prob{}{\hat s_+\,|\, s,a} = \Prob{}{s_+\,|\, s,a}$, then the stage cost defined in~\eqref{eq:hatL_def} satisfies $\hat L(s,a) = L(s,a)$, $\forall \, s\in \mathcal{S}$, with $\mathcal{S}$ defined in~\eqref{eq:ZeAssumption}.
It is interesting to discuss here the case $N\rightarrow \infty$ for which, under some conditions, the terminal cost can be dismissed. We detail this in the following Corollary.
\vspace{0.25cm}
\begin{Corollary} 
	\label{cor:ZeCorollary}
	Under the assumptions of Theorem \ref{thm:ZeTheorem} and under the additional assumption:
\begin{align}
\lim_{N\rightarrow\infty}\,  \E{}{\gamma^NV_\star(\hat s_N^\pi)} = 0, \label{eq:StabilityAssumption}
\end{align}
then 
\begin{align}
\hat V_\infty(s) = \lim_{N\rightarrow\infty}\, \min_{\pi}\E{}{\sum_{k=0}^{N-1}\, \gamma^k \hat L(\hat s^{ \pi}_k, \pi(\hat s^{ \pi}_k))} = V_\star(s), \label{eq:VhatInfinite}
\end{align}
and the equalities $\hat\pi(s) = \pi_\star(s)$ and $\hat Q_\infty(s,a) = Q_\star(s,a)$ hold for $\left |\,\mathbb{E}\left[V_\star\left(\hat s_+\right)\,|\, s,a\right] \,\right | < \infty$.
\end{Corollary}
\vspace{0.25cm}
\begin{IEEEproof}
Let us consider the $N$-step value function $\hat V_N^\pi$ associated to the stage cost $\hat L$, the state transition model $\Prob{}{\hat s_+\,|\, s,a}$ and a policy $\pi$ defined as:
\begin{subequations}
\label{eq:Vhatinf}
	\begin{align}
		\hspace{-1em}\hat V^\pi_\infty(s) &=\lim_{N\rightarrow\infty}\E{}{\sum_{k=0}^{N-1} \gamma^{k} \hat L(\hat s_k^\pi,\pi(\hat s_k^\pi)) } \label{eq:Vhat:Definf}\\
		&=Q_\star(s,\pi(s)) + \lim_{N\rightarrow\infty}\,\mathbb{E}\Bigg [-\gamma^NV_\star(\hat s_N^\pi)\nonumber \\
		& \hspace{9em}+ \sum_{k=1}^{N-1} \gamma^k A_\star(\hat s_k^\pi,\pi(\hat s_k^\pi))  \Bigg ], \label{eq:Vhat:Telescopicinf}
	\end{align}
	\end{subequations}
where \eqref{eq:Vhat:Telescopicinf} holds as, by assumption, all terms in the summations in \eqref{eq:Vhatinf} have finite expected values. This entails that
\begin{align}
&\hat V^{\pi_\star}_\infty(s)=\lim_{N\rightarrow\infty}\,V_\star(s) + \mathbb{E}\Bigg [-\gamma^NV_\star(\hat s_N^{\pi_\star}) \Bigg ].
\end{align}
Using \eqref{eq:StabilityAssumption}, we finally observe
that the following inequalities hold on $\mathcal{S}$:
\begin{align}
\hat V_\infty (s) = \min_\pi \hat V^\pi_\infty(s) = \hat V^{\pi_\star}_\infty(s) =V_\star(s), 
\end{align}
and are justified similarly to \eqref{eq:TrickyEqualitiesInProof}.

\end{IEEEproof}
Let us make next a few observations regarding Theorem \ref{thm:ZeTheorem}.
\begin{itemize}
\item Assumption \eqref{eq:ZeAssumption} requires that the model trajectories under policy $\pi_\star$ are contained within the set where the value function $V_\star$ is finite with a unitary probability. 
\item Assumption \eqref{eq:StabilityAssumption} can be construed as some form of stability condition on the model dynamics under policy $\pi_\star$. This observation is especially clear in the case $\gamma=1$, which then imposes the condition $\lim_{N\rightarrow\infty}\,\E{}{V_\star(\hat s_N^\pi)} = 0 $
\item Condition \eqref{eq:StabilityAssumption} is not required in Theorem \ref{thm:ZeTheorem}, however, it is highly desirable to fulfil it in practice when a finite-horizon is required, so that the terminal cost in \eqref{eq:VhatN:Definition} has a limited impact on the optimal policy $\hat\pi$.
\item Though very similar, Theorem~\ref{thm:ZeTheorem} and Corollary~\ref{cor:ZeCorollary} cover different cases, since taking the limit for $N\rightarrow\infty$ in Theorem~\ref{thm:ZeTheorem} does not require Assumption \eqref{eq:StabilityAssumption} to hold.
\item 
Theorem \ref{thm:ZeTheorem} proposes a modified stage cost and a terminal cost such that the finite-horizon problem \eqref{eq:VhatN:Definition} delivers the optimal policy $\pi_\star$. It ought to be noted that problem \eqref{eq:VhatN:Definition} would actually use $\pi_\star(\hat s_k)$ to select the inputs at every stage $k$ of the prediction $\hat s_{0,\ldots,N-1}$. However, identities (i)-(iii) of Theorem \ref{thm:ZeTheorem} do not necessarily require this restriction, i.e. it is sufficient to find a stage cost and a terminal cost such that the resulting optimal control problem yields $\pi_\star$ as its \textit{initial} policy (at stage $k=0$) only to get identities (i)-(iii). 
\end{itemize}

\section{ENMPC as a Function Approximator} \label{sec:NMPCusingRL}
We will now detail how the theory presented above applies to using a parametrized ENMPC scheme to approximate the optimal policy and value functions $\pi_\star$ and $V_\star,\,Q_\star$, even if the model underlying the ENMPC scheme is not highly accurate. We will detail in Section \ref{eq:RLENMPC} how the ENMPC parameters can be adjusted to achieve this approximation.

The constraints \eqref{eq:FiniteL} can be used in the ENMPC scheme to explicitly exclude undesirable states and inputs. Since the ENMPC scheme will be based on an imperfect model $f_\theta$, it will seek the minimization of the modified stage cost $\hat L(s,a)$ rather than the original one $L(s,a)$. As a result, while the pure input constraints $g(a) \leq 0$ are arguably fixed, the mixed constraints used in the ENMPC scheme ought to be modified as well, in order to capture the domain where $\hat L(s,a)$ is finite. Since $\hat L$ is not known a priori, the domain where it is finite depends, among other things, on the discrepancy between $f_\theta$ and \eqref{eq:TrueDynamics}, and will have to be learned. We will therefore consider introducing the parametrized mixed constraints $h_\theta\left(s,a\right) \leq 0$ in the ENMPC scheme, where $\theta$ will be parameters that can be adjusted via RL tools. Equation \eqref{eq:ENMPC:Perfect:Parametrization} and Corollary \ref{Cor:PerfectParam} below provide a more formal explanation of these observations.

Since RL tools cannot handle infinite penalties, we will need to consider a relaxed version of $L$ and of the mixed constraints $h_\theta$. We can now formulate the parametrized ENMPC scheme that will serve as a function approximation in the RL tools.

We will consider a parametrization of the value function $V_\star$ using the following ENMPC scheme parametrized by $\theta$:
\begin{subequations}
	\label{eq:param_nmpc:valuefunction}
\begin{align}
V_\theta(s) =  \min_{u,x,\sigma}\ \ & \lambda_\theta(x_0) + \gamma^N \left(V^\mathrm{f}_\theta(x_N) + w_\mathrm{f}^\top \sigma_N\right)
 \nonumber \\
&\hspace{2em}+ \sum_{k=0}^{N-1} \gamma^k \left (l_\theta(x_k,u_k) + w^\top \sigma_k \right ) \label{eq:param_nmpc:cost}\\
\mathrm{s.t.} \ \ & x_{k+1} = f_\theta\left(x_k,u_k\right),\quad x_0 = s, \label{eq:param_nmpc:dynamics}\\
& g\left(u_k\right) \leq 0, \label{eq:param_nmpc:input_const} \\
& h_\theta\left(x_k,u_k\right) \leq \sigma_k,\quad h^\mathrm{f}_\theta(x_N) \leq \sigma_N. \label{eq:Const:Relaxation} 
\end{align}
\end{subequations}
Problem \eqref{eq:param_nmpc:valuefunction} is a classic ENMPC formulation when $\gamma=1$ and $\lambda_\theta = 0$ \cite{Rawlings2009b,Grune2011}. Note that we have used $x,u$ in order to clearly distinguish the ENMPC prediction from the actual closed-loop state and control trajectory. 

We observe that the ENMPC scheme \eqref{eq:param_nmpc:valuefunction} holds a model parametrization $f_\theta$, a constraint parametrization $h_\theta$ as discussed above, and a parametrization of the stage cost $l_\theta$ and terminal cost $V_\theta^\mathrm{f}$. The extra cost $\lambda_\theta$ is discussed in detail in Section \eqref{sec:RotateMyAss}. Reasonable guesses for these functions are to use $l_\theta = l$, $h_\theta = h$, and any classic heuristic to build the terminal cost approximation $V_\theta^\mathrm{f}$, such as e.g. a quadratic cost stemming from the LQR approximation of the ENMPC scheme. RL tools are then used to modify these initial guesses towards higher closed-loop performances.

The $\ell_1$ relaxation of the mixed constraints \eqref{eq:Const:Relaxation} relying on slack variables $\sigma_k$ is fairly standard in practical implementation of (EN)MPC schemes. For $w,w_\mathrm{f}$ large enough, the solution to \eqref{eq:param_nmpc:valuefunction} is identical to the unrelaxed one whenever a feasible trajectory exists for the initial state $s$~\cite{Scokaert1999a}. In this case, we refer to the relaxation as \emph{exact}. The specific role of the relaxation will be detailed in Section \ref{eq:RL:ConstraintsRelaxation}.  Function $\lambda_\theta$ in \eqref{eq:param_nmpc:cost} is not required anywhere in the following developments, but will play a central role in forming nominal stability guarantees of the ENMPC scheme \eqref{eq:param_nmpc:valuefunction}, see Section~\ref{sec:ENMPC}. We define the policy:
\begin{align}
\pi_\theta(s) = u_0^\star, \label{eq:ENMPC:Policy}
\end{align}
where  $u_0^\star$ is the first element of the input sequence $u_0^\star,\ldots, u_{N-1}^\star$ solution of \eqref{eq:param_nmpc:valuefunction} for a given $s$. We additionally define the action-value function $Q_\theta$:
\begin{subequations}
	\label{eq:param_nmpc}
\begin{align}
Q_\theta(s,a) = \min_{u,x}&\quad \eqref{eq:param_nmpc:cost} \\
\mathrm{s.t.} &\quad \eqref{eq:param_nmpc:dynamics}-\eqref{eq:Const:Relaxation}, \label{eq:constrFromV}\\
&\quad u_0 = a. \label{eq:Input:Embedding}
\end{align}
\end{subequations}
Note that the proposed parametrization trivially satisfies the fundamental equalities underlying the Bellman equations, i.e.:
\begin{align}
\pi_\theta(s) = \mathrm{arg}\min_a\, Q_\theta(s,a),\quad V_\theta(s) = \min_a\, Q_\theta(s,a).
\end{align}

Let us make some key observations on \eqref{eq:param_nmpc:valuefunction}-\eqref{eq:ENMPC:Policy}. The deterministic model \eqref{eq:param_nmpc:dynamics} can be construed as a special case of the stochastic state transition \eqref{eq:ModelDynamics}, using:
\begin{align}
\Prob{}{\hat s_+\,|\, s,a} &= \delta(\hat s_+ - f\left(s,a\right)),
\end{align}
where $\delta$ is the Dirac distribution. Moreover, suppose that we can select $\theta$ such that $\lambda_\theta=0$ and such that the stage cost and constraints in  \eqref{eq:param_nmpc:valuefunction} satisfy:
\begin{subequations}
\label{eq:ENMPC:Perfect:Parametrization}
\begin{align}
\hat L\left(s,a\right) &= l_\theta \left(s,a\right) +\mathcal{I}_\infty\left(h_{\theta}\left(s,a\right)\right)+ \mathcal{I}_\infty\left(g_{\theta}\left(a\right)\right), \\
V_\star\left(s\right) &= V_{\theta}^\mathrm{f}(s) + \mathcal{I}_\infty\left(h_{\theta}^\mathrm{f}(s)\right),
\end{align}
\end{subequations}
for $\hat L$ given by \eqref{eq:hatL_def}, where $l_{\theta},\,V_{\theta}^\mathrm{f}<\infty$. 
\begin{Corollary}[of Theorem~\ref{thm:ZeTheorem}] \label{Cor:PerfectParam}
	Assume that the NMPC scheme~\eqref{eq:param_nmpc:valuefunction} is parametrized using a rich enough parametrization with an exact relaxation (i.e. $w,w_\mathrm{f}$ large enough). Then, the NMPC scheme~\eqref{eq:param_nmpc:valuefunction} delivers the optimal policy $\pi_\star$ and value functions $V_\star,\,Q_\star$ for any state $s$ for which assumption \eqref{eq:ZeAssumption} is satisfied.
\end{Corollary}
\begin{IEEEproof}
	By assumption, there exists $\theta$ such that~\eqref{eq:ENMPC:Perfect:Parametrization} holds, then Theorem \ref{thm:ZeTheorem} directly yields the desired result.
\end{IEEEproof}
Note that assumption \eqref{eq:ZeAssumption} entails that $\mathcal{S}$ is the forward-invariant set of the dynamics \eqref{eq:param_nmpc:dynamics} under the policy $\pi_\star$, associated to the condition $|V_\star(s)| < \infty$. 


Unfortunately, a parameter $\theta$ satisfying \eqref{eq:ENMPC:Perfect:Parametrization} is clearly not guaranteed to exist, and does not exist in most non-trivial practical cases. Indeed, the modified stage cost $\hat L$ defined by  \eqref{eq:hatL_def} can be highly intricate, and satisfying \eqref{eq:ENMPC:Perfect:Parametrization} can require a very elaborate parametrization of the ENMPC cost and constraints. Even if a $\theta$ satisfying \eqref{eq:ENMPC:Perfect:Parametrization} does exist, finding it is arguably highly difficult as evaluating \eqref{eq:hatL_def} can be extremely demanding and requires the knowledge of the real stochastic state transition \eqref{eq:ModelDynamics}. In this paper, we propose to circumvent this difficulty by (i) relying on a limited parametrization of the ENMPC scheme, at the price of not achieving $\pi_\theta = \pi_\star$ exactly, and (ii) deploying RL techniques in order to adjust the ENMPC parameters $\theta$, so as to avoid computing $\hat L$ altogether, see Section \ref{eq:RLENMPC}. In the RL context, we ought to consider the ENMPC scheme \eqref{eq:param_nmpc:valuefunction} as a function approximator for $V_\star,\, Q_\star$ and $\pi_\star$.

The choice of cost parametrization is clearly important, but beyond the scope of this paper. In the example below, see Section \ref{sec:Examples}, we have made fairly trivial choices and obtained interesting results nonetheless. It is clear, however, that a rich parametrization is in theory desirable. It is also desirable that the stage cost and terminal cost approximations, $l_\theta$ and $V^\mathrm{f}_\theta$, always remain (quasi-)convex in order to facilitate the computation of the NMPC solution. In contrast, the parametrization of the initial cost $\lambda_\theta$ can be unrestricted. In the future, we will investigate rich function approximations, including e.g. positive sums of convex functions and sum-of-squares approaches.

\subsection{Robust NMPC Using Scenario Trees} \label{sec:Tree}
Robust NMPC implemented via scenario tree approaches also readily fits in the framework presented here. Indeed, the scenario tree can be construed as a stochastic process with a discrete probability distribution. More specifically, the stochastic model state transition for a scenario tree dynamics reads as:
\begin{align}
\Prob{}{\hat s_+\,|\, s,a} = \sum_{i=1}^{N_{\mathrm{s}}} W_i \delta\left(\hat s_+ -  f_i(s,a)\right),
\end{align}
where  $f_{1,\ldots,N_{\mathrm{s}}}$ are the $N_{\mathrm{s}}$ different dynamics underlying the scenario tree, and $W_{1,\ldots,N_{\mathrm{s}}}$ are the associated probabilities, with $W_i\geq 0$ and $\sum_{i=1}^{N_{\mathrm{s}}} W_i=1$. All the observations made in Section \ref{sec:NMPCusingRL} are then also valid in this context. One ought then to see the discrete probability distribution $w_i$ as part of the parameters to be adjusted in the (E)NMPC scheme. 

\subsection{Model Parametrization} \label{sec:ModelAdaptation}

As detailed in Section \ref{sec:NMPCusingRL} the ENMPC scheme \eqref{eq:param_nmpc:valuefunction} can in principle capture the optimal policy $\pi_\star$ without having to adjust the model \eqref{eq:param_nmpc:dynamics}. This observation, however, does not preclude an adaptation of the model \eqref{eq:param_nmpc:dynamics} in order to drive the NMPC policy $\pi_{\theta}$ towards the optimal one $\pi_\star$, and one can easily argue that allowing such an adaptation introduces additional freedom for the NMPC scheme  \eqref{eq:param_nmpc:valuefunction} to better approximate the optimal policy $\pi_\star$. 
The interplay between the cost and constraints adaptation and the model adaptation is the object of current research. 




\section{Reinforcement-Learning for ENMPC} \label{eq:RLENMPC}

Theorem \ref{thm:ZeTheorem} guarantees that it is in theory possible to generate the optimal policy and value functions using an ENMPC scheme based on a possibly inaccurate model. 
In practice, one has to rely on ad-hoc parametrization $\theta$ of the NMPC scheme, yielding the value function $Q_\theta(s,a),\,V_\theta(s)$ and the policy $\pi_\theta(s)$. The goal is then to adjust the parameters $\theta$ such that the ENMPC scheme policy fits the optimal policy as closely as possible. 

Because computing $\hat L$ given by \eqref{eq:hatL_def} is difficult and requires the knowledge of the true dynamics, we will rely on RL techniques to adjust the parameters $\theta$. We will focus here on classical RL approaches. These techniques typically require the sensitivities of the value functions. We briefly detail next how to compute these sensitivities for \eqref{eq:param_nmpc:valuefunction}, \eqref{eq:ENMPC:Policy}, and \eqref{eq:param_nmpc}.

\subsection{Sensitivities of the ENMPC scheme} \label{sec:Qlearning}
We detail next how to evaluate the gradients of functions $Q_\theta,\,V_\theta,\, \pi_\theta$. 
To that end, let us define the Lagrange function associated to the ENMPC Problem~\eqref{eq:param_nmpc} as
\begin{subequations}
	\begin{align*}
	\mathcal{L}_\theta(s,y) = \ & \lambda_\theta(x_0)+\gamma^N V^\mathrm{f}_\theta(x_N)  + \chi_0^\top\left(x_0 - s\right) + \mu_N^\top h^\mathrm{f}_\theta(x_N)\\ 
	& + \sum_{k=0}^{N-1}  \chi_{k+1}^\top\left(f_\theta\left(x_k,u_k\right) - x_{k+1} \right) + \nu_k^\top g_\theta\left(u_k\right)\\
	&+\gamma^k L_\theta(x_k,u_k) + \mu_{k}^\top h_\theta\left(x_k,u_k\right) + \zeta^\top(u_0-a),
	\end{align*}
\end{subequations}
where $\chi,\mu,\nu,\zeta$ are the multipliers associated to constraints \eqref{eq:param_nmpc:dynamics}-\eqref{eq:Const:Relaxation} and 
\eqref{eq:constrFromV}-\eqref{eq:Input:Embedding} respectively, and we will note $y=(x,u,\chi,\mu,\nu,\zeta)$ the primal-dual variables associated to \eqref{eq:param_nmpc}. 

Note that, for $\zeta=0$, $\mathcal{L}_\theta(s,y)$ is the Lagrange function associated to the NMPC problem~\eqref{eq:param_nmpc:valuefunction}.
We observe that \cite{Buskens2001}
\begin{align}
\label{eq:NMPCQgradient}
\nabla_\theta Q_\theta(s,a) = \nabla_\theta \mathcal{L}_\theta(s,y^\star)
\end{align}
holds for $y^\star$ given by the primal-dual solution of \eqref{eq:param_nmpc}. The gradient \eqref{eq:NMPCQgradient} is therefore straightforward to build as a by-product of solving the NMPC problem \eqref{eq:param_nmpc}. We additionally observe that
\begin{align}
\label{eq:NMPCVgradient}
\nabla_\theta\,V_\theta(s) = \nabla_\theta \mathcal{L}(s,y^\star),
\end{align}
for $y^\star$ given by the primal-dual solution to \eqref{eq:param_nmpc:valuefunction} completed with $\zeta^\star=0$. Finally, the gradient of the NMPC policy with respect to the parameters $\theta$ is given by \cite{Buskens2001}:
\begin{align}
\label{eq:dpolicy_dtheta}
\nabla_{\theta}\pi_\theta(s) = - \nabla_{\theta } \xi_\theta(s,y^\star)\nabla_{y} \xi_\theta(s,y^\star)^{-1} \frac{\partial y}{\partial u_0},
\end{align}
for $y^\star$ given by the primal-dual solution to \eqref{eq:param_nmpc:valuefunction} with $\zeta^\star=0$, and where $\xi_\theta(s,y)$ gathers the primal-dual KKT conditions underlying the NMPC scheme \eqref{eq:param_nmpc:valuefunction}.

We remark that~\eqref{eq:NMPCQgradient} and~\eqref{eq:NMPCVgradient} are well-defined for any $s,\theta$ such that no inequality constraint in the ENMPC schemes \eqref{eq:param_nmpc:valuefunction} and \eqref{eq:param_nmpc}, respectively, is weakly active. When an inequality constraint is weakly active, \eqref{eq:NMPCQgradient} and~\eqref{eq:NMPCVgradient} may be defined only up to the subgradients generated by the possible active sets. This technical issue is fairly straightforward to circumvent in practice by use of interior-point techniques when solving \eqref{eq:param_nmpc:valuefunction} and \eqref{eq:param_nmpc}. Additionally, the policy gradient~\eqref{eq:dpolicy_dtheta} is only valid if the ENMPC~\eqref{eq:param_nmpc:valuefunction} satisfies the linear independence constraint qualification, and the second-order sufficient conditions~\cite{Buskens2001}, which are typically satisfied by properly formulated ENMPC schemes.

We detail next how TD-learning is deployed on the NMPC scheme \eqref{eq:param_nmpc}, both in an on- and off-policy fashion.

\subsection{$Q$-learning for (E)NMPC}
A classical approach to $Q$-learning~\cite{Sutton1998} is based on parameter updates driven by the temporal-difference with instantaneous policy updates
\begin{subequations}
	\label{eq:Qlearning:OnPolicy}
	\begin{align}
	\tau_k &= L_\theta(s_k,a_k) + \gamma\,V_\theta(s_{k+1}) - Q_\theta(s_k,a_k), \label{eq:TDError:OnPolicy}\\
	\theta &\leftarrow \theta + \alpha\tau_k\nabla_\theta Q_\theta(s_k,a_k), \label{eq:Qlearning:OnPolicy:Update}
	\end{align}
\end{subequations}
where
\begin{align}
L_\theta(s_k,a_k) = l_\theta(x_k,a_k) + w^\top \max (0,h_\theta\left(x_k,a_k\right) ),
\end{align}
and where $a_k$ is selected according to the NMPC policy $\pi_\theta(s)$, with the possible addition of occasional random exploratory moves \cite{Sutton1998}. The scalar $\alpha>0$ is a step-size commonly used in stochastic gradient-based approaches. A version of $Q$-learning with batch policy updates reads as:
\begin{subequations}
	\label{eq:Qlearning:OffPolicy}
	\begin{align}
	\tau_k &= L_{\tilde\theta}(s_k,a_k) + \gamma\,V_{\tilde\theta}(s_{k+1}) - Q_{\tilde\theta}(s_k,a_k), \\
	\tilde \theta &\leftarrow \tilde \theta + \alpha\tau_k\nabla_{\tilde\theta} Q_{\tilde\theta}(s_k,a_k),
	\end{align}
\end{subequations}
where $a_k$ is selected according to the fixed NMPC policy $\pi_\theta(s)$ while the learning is performed within an alternative NMPC scheme based on the parameters $\tilde\theta$, and not applied to the real system. Note that the on-policy approach \eqref{eq:Qlearning:OnPolicy} entails that changes in the NMPC parameters $\theta$ are readily applied in closed-loop after each update \eqref{eq:Qlearning:OnPolicy:Update}, while the off-policy approach \eqref{eq:Qlearning:OffPolicy} allows one to learn a new set of policy parameters $\tilde\theta$ while deploying the original NMPC scheme, based on the parameters $\theta$, on the real system. The learned parameters $\tilde\theta$ can then be introduced in closed-loop at convenience, by performing the replacement $\theta\leftarrow\tilde\theta$, e.g. after they have converged and after a formal verification of the corresponding NMPC scheme has been carried out, see e.g. \cite{Simon2016,Lofberg2012}. 

It ought to be made clear here that RL methods of the type \eqref{eq:Qlearning:OnPolicy} or \eqref{eq:Qlearning:OffPolicy} yield no guarantee to find the global optimum of the parameters. 
This limitation pertains to most applications of RL relying on nonlinear function approximators such as the commonly used DNN. In practice, however, RL improves the closed-loop performance over the one of the initial parameters.


\subsection{Role of the Constraints Relaxation} \label{eq:RL:ConstraintsRelaxation}


We can now further discuss the $\ell_1$ constraints relaxation \eqref{eq:Const:Relaxation} in the light of the TD approaches \eqref{eq:Qlearning:OnPolicy} and \eqref{eq:Qlearning:OffPolicy}. In the absence of constraints relaxation, the value functions take infinite values when a constraint violation occurs, and they are therefore meaningless in the context of RL, as only finite value functions can be used in \eqref{eq:Qlearning:OnPolicy}-\eqref{eq:Qlearning:OffPolicy}. The proposed constraint relaxation ensures that the value functions retain finite value even in the presence of constraints violation, such that the RL updates  \eqref{eq:Qlearning:OnPolicy}-\eqref{eq:Qlearning:OffPolicy} remain well-defined and meaningful. This technical observation has a fairly simple and generic interpretation: any form of learning is meaningless if infinite penalties are assigned to violating limitations.

In practice, violating some safety-critical constraints may be unacceptable. In that context, avoiding the violation of crucial constraints ought to be prevented from the formulation of the (E)NMPC scheme. Here, robust NMPC techniques are arguably an important tool to avoid such difficulties. The applicability of the proposed theory to the robust (E)NMPC formulations of Sec. \ref{sec:Tree} is therefore of crucial importance for safety-critical applications. The interplay of robust (E)NMPC with RL and the handling of safety-critical constraints will be the object of future publications.


\subsection{Deterministic Policy Gradient Methods for ENMPC}
It is useful to underline here that $Q$-learning techniques seek the fitting of $Q_\theta$ to $Q_\star$ under some norm, with the hope that $Q_\theta\approx Q_\star$ will result in $\pi_\theta \approx \pi_\star$. There is, however, no a priori guarantee that the latter approximation holds when the former does. In order to formally maximize the performance of policy $\pi_\theta$, it is useful to turn to policy gradient methods. For the sake of brevity we propose to focus on deterministic policy gradient methods here \cite{Silver2014}, based on the policy gradient equation:
\begin{align}
\nabla_\theta J(\pi_\theta) = \mathbb{E}{}\left[\nabla_\theta  \pi_\theta(s) \nabla_a Q_{\pi_\theta}(s,a) \right], \label{eq:PiGrad}
\end{align}
where $J$ is the expected closed-loop cost associated to running policy ${\pi_\theta}$ on the real system, and $Q_{\pi_\theta}$ the corresponding action-value function. Note that the expectation $\mathbb{E}$  is taken over trajectories of the real system subject to policy $\pi_\theta$. A necessary condition of optimality for policy $\pi_\theta$ is then:
\begin{align}
\nabla_\theta J(\pi_\theta) =  0. \label{eq:PiOptimality}
\end{align}
Deterministic policy gradient methods are often built around the TD actor-critic approach, based on \cite{Silver2014}:
\begin{align}
&\tau_k =  L(s_k,a_k) + \gamma\,Q_{w}(s_{k+1},\pi_\theta(s_{k+1})) - Q_{w}(s_k,a_k)\nonumber \\
&w\leftarrow w + \alpha_w \tau_k  \nabla_w Q_{w}(s_k,a_k)\nonumber\\
&\theta \leftarrow \theta + \alpha_\theta \nabla_\theta  \pi_\theta(s_k) \nabla_a Q_{w}(s_k,\pi_\theta(s_{k})) 
\end{align}
for some $\alpha_w, \alpha_\theta > 0$ small enough, where $Q_w \approx Q_{\pi_\theta}$ is an approximation of the corresponding action-value function. Computationally efficient choices of action-value function parametrization $Q_w$ in the ENMPC context will be the object of future publications.

\section{RL and Stable Economic NMPC} \label{sec:ENMPC}

%


The main idea in economic NMPC is to optimize performance (defined by a suitably chosen cost) rather than penalizing deviations from a given reference. Since the cost is generic, the value function is not guaranteed to be positive-definite and proving stability becomes challenging.
The main idea for proving stability is to introduce a cost modification, called \emph{rotation}, which does not modify the optimal solution, but yields a positive-definite value function, and hence is a Lyapunov function, such that nominal stability follows.

Section~\ref{sec:ZeSection} focused on learning the optimal policy even in case a wrong model $\Prob{}{\hat s_+\,|\,s,u}$ is used. In this section, we aim at enforcing nominal stability even if the corresponding stage cost $\hat L$ is indefinite\footnote{We remark that in this context, stability is obtained for the model used by NMPC for predictions. Future work will investigate obtaining stability guarantees for the real process $\Prob{}{s_+\,|\,s,u}$.}. To this end, in the NMPC cost we replace $\hat L$ by a newly-defined stage cost $\bar L$, which we force to be positive-definite. We then use a cost modification in order to recover the correct, possibly indefinite, value and action-value functions corresponding to $\hat L$.
Throughout the section, we assume that the value and action-value functions are bounded on some set.




In order to show how the cost modification can be implemented by the term $\lambda_\theta$ in~\eqref{eq:param_nmpc:valuefunction}, we first introduce the proposed cost modification as a generalization of the standard one, prove that it does not modify the optimal policy and discuss how it is used to learn the optimal value and action-value functions. Then, we discuss how the cost modification relates to the cost rotation typically used in ENMPC. Finally, we illustrate how the cost modification can be used in a parametrized ENMPC scheme to be used as function approximator within RL.

\subsection{Generalized Cost Rotation} \label{sec:RotateMyAss}

Consider modifying the cost of the problem according to 
\begin{subequations}
	\label{eq:PiPreservation}
	\begin{align}
	\hspace{-0.8em}\bar L\left(s,a\right) &=  \hat L\left(s,a\right) +  \varLambda\left(s,a\right) - \gamma\E{}{\varLambda\left(\hat s_+,\hat\pi(\hat s_+)\right)\,|\, s,a}, \label{eq:PiPreservation:Condition:L1}\\
	\bar V_\mathrm{f}(s) &= V_\star (s) + \varLambda\left(s,\hat \pi(s)\right), \label{eq:PiPreservation:Condition:Vf}
	\end{align}
	where $\bar V_\mathrm{f}(s)$ denotes the rotated terminal cost and where
	\begin{align}
	\varLambda\left(s,a\right) &\geq \varLambda(s,\hat\pi(s)),\quad \,\forall\,s,a \label{eq:PiPreservation:Condition:L2}
	\end{align}
\end{subequations}
holds. Moreover, we require $\varLambda(s,a)$ to be such that $\bar L\left(s,a\right)$ is finite whenever $\hat L\left(s,a\right)$ is finite.

It is important to underline that defining a function $\varLambda(s,a)$ strictly satisfying condition~\eqref{eq:PiPreservation:Condition:L2} is in general hard, as it requires knowledge of $\hat \pi$. A simpler choice satisfying~\eqref{eq:PiPreservation:Condition:L2} with equality is $\varLambda(s,a)=\lambda(s)$, where $\lambda$ is any function satisfying the boundedness assumption on $\bar L(s,a)$.

\begin{theorem}
	\label{thm:policy_preserving_cost_modification}
	The modification~\eqref{eq:PiPreservation} 
	preserves the optimal control policy $\hat \pi$ corresponding to $\hat L$ and the model $\Prob{}{\hat s_+\,|\,s,u}$. Moreover, the optimal value and action-value functions satisfy
	\begin{subequations}
		\label{eq:PiPreservingModificationValues}
		\begin{align}
		\bar V_N(s) &= \hat V_N(s) + \varLambda(s,\hat \pi(s)), \label{eq:PiPreservingModificationValueV} \\
		\bar Q_N(s,a) &= \hat Q_N(s,a) + \varLambda(s,a).\label{eq:PiPreservingModificationValueQ}
		\end{align}
	\end{subequations}
\end{theorem}
\begin{IEEEproof} 
	We will first prove the theorem for $N=1$ and then proceed by induction.
	%
	%
	%
	%
	The definition of action value function reads
	\begin{align}
	\label{eq:hat_action_vf_def}
	\hat Q_1\left(s,a\right) = \hat L\left(s,a\right)  + \gamma\E{}{V_\star\left(\hat s_+\right)\,|\, s,a}.
	\end{align}
	Then, we can write
	\begin{align*}
	&\hat Q_1\left(s,a\right) + \varLambda\left(s,a\right)  \\
	&\hspace{4em}= \hat L\left(s,a\right) + \varLambda\left(s,a\right) + \gamma\E{}{V_\star\left(\hat s_+\right)\,|\, s,a} \\
	&\hspace{4em}= \bar L\left(s,a\right) + \gamma\E{}{\hat V_\mathrm{f}\left(\hat s_+\right)\,|\, s,a}= \bar Q_1\left(s,a\right).
	\end{align*}
	Since by definition $\hat Q_1\left(s,a\right) > \hat Q_1\left(s,\hat \pi(s_+)\right)$, $\forall \, a \notin \hat \pi(s_+)$ and by construction~\eqref{eq:PiPreservation:Condition:L2} holds, we obtain that
	\begin{align*}
	\bar Q_1\left(s,a\right) > \bar Q_1\left(s,\hat \pi(s_+)\right), \ \ \forall \, a \notin \hat \pi(s_+).
	\end{align*}
	Therefore, the optimal policy is preserved and
	\begin{align*}
	\bar V_1(s) = \hat V_1(s) + \varLambda(s,\hat \pi(s))= \hat V_\star(s) + \varLambda(s,\hat \pi(s)).
	\end{align*}
	
	We now use bootstrapping to obtain
	\begin{align*}
	&\hat Q_{k+1}\left(s,a\right) + \varLambda\left(s,a\right)  \\
	&\hspace{4em}= \hat L\left(s,a\right) + \varLambda\left(s,a\right) + \gamma\E{}{V_k\left(\hat s_+\right)\,|\, s,a} \\
	&\hspace{4em}= \bar L\left(s,a\right) + \gamma\E{}{\hat V_k\left(\hat s_+\right)\,|\, s,a}= \bar Q_{k+1}\left(s,a\right).
	\end{align*}
	Therefore, cost modification~\eqref{eq:PiPreservation} preserves the policy over any horizon $N$, i.e. $\bar \pi=\hat \pi$, and~\eqref{eq:PiPreservingModificationValues} hold.
	%
\end{IEEEproof}



We will show next that, for any given policy, the modified NMPC scheme can yield any bounded value function, while this does not hold for the action-value function.
Consequently any attempt at learning the optimal policy by solely relying on learning the value function and the proposed modification cannot succeed.


Consider a problem formulated using a stage cost $\check L(s,a)$, with the corresponding value functions associated to the optimal policy $\check\pi$
\begin{subequations}
	\begin{align}
	\check V_N(s)&=\check L\left(s,\check \pi(s)\right) +\gamma\E{}{\check V_{N-1}\left(\check s_+\right)\,|\,s,\check\pi(s)}, \\
	\check Q_N(s,a) &= \check L\left(s,a\right) +\gamma\E{}{\check V_{N-1}\left(\check s_+\right)\,|\,s,a}.
	\end{align}
\end{subequations}
\begin{Lemma}
	\label{lem:value_function_rotation}
	Consider the set $\check{\mathcal{S}}$ such that for all $s\in\check{\mathcal{S}}$ both $\check V_N(s)$ and $\hat V_N(s)$ are bounded. Then, there exists a cost modification $\Lambda(s,a)=\lambda(s)$, $\forall \, a$ such that 
	\begin{align}
	 \check V_N(s) + \lambda(s) = \hat V_N(s).
	\end{align}
\end{Lemma}
\begin{IEEEproof}
	The result is obtained by choosing $\lambda(s)= \hat V_N(s)- \check V_N(s)$.
\end{IEEEproof}

\subsection{Strict Dissipativity}

Since most of the literature on dissipativity-based ENMPC focuses on deterministic formulations, in this subsection we also restrict to the deterministic case and we adopt $\gamma=1$, as is usual in the literature on ENMPC.
Strict dissipativity is typically used in ENMPC in order to obtain a positive-definite cost to construct a Lyapunov function and prove closed-loop stability. 
Strict dissipativity holds if there exists a function $\lambda$ such that 
\begin{align}
\label{eq:strict_diss}
\lambda(s_+) - \lambda(s)  \leq - \rho(\| s-s_\mathrm{e}\|) + L(s,a) - L(s_\mathrm{e},a_\mathrm{e}),
\end{align}
with $\rho$ a positive-definite function and
\begin{align*}
(s_\mathrm{e},a_\mathrm{e}) = \underset{s,a}{\mathrm{argmin}} \ \ L(s,a) && \mathrm{s.t.} \ \ s=f(s,a).
\end{align*}
In~\cite{Amrit2011a} it has been proven that strict dissipativity is sufficient for closed-loop stability, provided that $\lambda$ is bounded and the constraint set is compact. In~\cite{Mueller2015} it has been proven that, under a controllability assumption and if $(s_\mathrm{e},a_\mathrm{e})$ lies in the interior of the constraint set, strict dissipativity with a bounded function $\lambda$ is also necessary for closed-loop stability on a compact constraint set.

For simplicity and without loss of generality, we assume that $L(s_\mathrm{e},a_\mathrm{e})=0$. 
\begin{Lemma}
	\label{lem:pd_modification}
	If there exists a function $\Lambda(s,a)$ such that $\min_a \Lambda(s,a)=\Lambda(s,\hat \pi(s))=\lambda(s)$, with $\lambda(s)$ satisfying the strict dissipativity inequality~\eqref{eq:strict_diss}, then $\bar L(s,a) \geq \rho(\|s-s_\mathrm{e}\|)\geq 0$ holds.
\end{Lemma}
\begin{IEEEproof}
	The proof is obtained by noting that
	\begin{align*}
	\bar L(s,a) &= L(s,a) + \Lambda(s,a) - \Lambda(\hat s_+,\hat \pi(s_+)) \\
	&\geq L(s,a) + \lambda(s) - \lambda(\hat s_+) \geq \rho(\| s-s_\mathrm{e}\|),
	\end{align*}
	where the first inequality follows from~\eqref{eq:PiPreservation:Condition:L2}, 
	and the second inequality is a direct consequence of~\eqref{eq:strict_diss} and $L(s_\mathrm{e},a_\mathrm{e})=0$.
\end{IEEEproof}
Lemma~\ref{lem:pd_modification} is of paramount importance in the context of ENMPC because it establishes that any stabilizing optimal policy $\pi_\star$ originating from any given stage cost $L(s,a)$ can be learned by using a parametrization that yields a positive-definite stage cost $\bar L(s,a)$. 
Note that the proposed cost modification is a generalization of the cost rotation used for ENMPC, as we replace $\lambda(s)$ with $\Lambda(s,a)$.

We ought to stress here that though the results presented in this section pertain to the deterministic case, we expect them to extend to the stochastic case as well. Unfortunately a mature dissipativity theory for stochastic ENMPC has not been developed yet. It would arguably not be surprising if a stochastic dissipativity criterion can be put in the form
\begin{align*}
\bar L(s,a) 
& \geq \rho(\| s-s_\mathrm{e}\|),
\end{align*}
with $\bar L$ defined in~\eqref{eq:PiPreservation} in combination with appropriate terminal conditions, hence offering a generalization of \cite{Sopasakis2017}. These questions will be the object of future reasearch.

\subsection{Economic Reward and NMPC Parametrization}

In the following, we will first show that the proposed cost modification can be reduced to a cost on the initial state, which is introduced in the ENMPC scheme~\eqref{eq:param_nmpc:valuefunction} as the term $\lambda_\theta$. Second, we will explain how we can use a stability-enforcing positive-definite stage cost to approximate the value and action-value function of the economic cost.


For simplicity, we use the cost modification $\varLambda(s,a)=\lambda(s)$. Note that we drop the dependence on the control in $\varLambda$ for practical reasons: (a) keeping it would require knowledge of the optimal policy and (b) $\lambda(s)$ is a valid choice in the sense that it satisfies~\eqref{eq:PiPreservation:Condition:L2}. 

We observe that the cost modification can be summarized by the initial cost $\lambda(x_0)$ by relying on the fact that for all predicted trajectories the modified cost reads as~\cite{Diehl2011,Amrit2011a}
\begin{align*}
&\phantom{= \,}\gamma^{N}V^\mathrm{f}(x_N) + \sum_{k=0}^{N-1} \gamma^{k}l(x_k,u_k) + \gamma^k \lambda(x_k) - \gamma^{k+1}\lambda(x_{k+1})\\
&=\lambda(x_0)+\gamma^{N}V^\mathrm{f}(x_N) + \sum_{k=0}^{N-1} \gamma^{k}l(x_k,u_k).
\end{align*}
In~\eqref{eq:param_nmpc:valuefunction}, we parametrize the cost modification as $\lambda_\theta(x_0)$

We now turn to the question of enforcing stability in the parametrized ENMPC scheme. 
In the literature on EMPC the cost modification is used to prove stability in case the stage cost is not positive-definite. 
In our case we are interested in the opposite: we would like to use a positive-definite stage cost to enforce stability while the cost modification is used in order to obtain value and action-value functions corresponding to the economic (non positive-definite) cost.

Thanks to Theorem~\ref{thm:ZeTheorem} and Lemma \ref{lem:pd_modification}, introducing the term $\lambda_\theta$ in \eqref{eq:param_nmpc:valuefunction} guarantees that if the optimal policy $\pi_\star$ is stabilizing for the model dynamics \eqref{eq:param_nmpc:dynamics}, then the ENMPC scheme \eqref{eq:param_nmpc:valuefunction} can deliver the optimal policy $\pi_\star$ and value functions $V_\star,\,Q_\star$ with a  stage cost $l_\theta$ lower-bounded by $\mathcal{K}_\infty$.

\section{Analytical Case Study: The LQR Case} \label{sec:LQR}

One of the few cases where \eqref{eq:ENMPC:Perfect:Parametrization} can be constructed and satisfied exactly is the LQR case. We use it as a simple illustration of the meaning of $\hat L$, Assumption~\eqref{eq:StabilityAssumption} and the cost modification $\varLambda(s,a)$.
Consider a centered linear-quadratic-gaussian control problem with the true dynamics and stage cost:
\begin{align}
\label{eq:lqr_dynamics}
s_+ &= As+Ba + e,\quad e\sim \mathcal{N}\left(0,\Sigma\right), \\
L\left(s,a\right) &= \tightMatr{c}{s \\ a}^\top \tightMatr{cc}{T & N \\ \star & R}\tightMatr{c}{s \\ a}.
\end{align}
The associated value functions if they exist read as:
\begin{align}
V_\star &= s^\top Ss + V_0,\\
Q_\star &= \tightMatr{c}{s \\ a}^\top \tightMatr{cc}{T + \gamma A^\top SA & N + \gamma A^\top SB \\ \star & R + \gamma B^\top S B}\tightMatr{c}{s \\ a} + V_0,\nonumber
\end{align}
where $V_0 = \gamma\left(1-\gamma\right)^{-1}\mathrm{Tr}\left(S\Sigma\right)$. 
Matrix $S$ and the associated optimal policy $K_\star$ are delivered by the Schur complement of the quadratic form in $Q_\star$, i.e. the discounted LQR equations:
\begin{subequations}
\label{eq:starSstarK}
\begin{align}
T + \gamma A^\top S A  &= S + \left(N + \gamma A^\top S B \right)K_\star,  \\
\left(R + \gamma B^\top S B \right)K_\star &= N^\top + \gamma B^\top S A.  \label{eq:lqr_optimal_feedback}
\end{align}
\end{subequations}

\subsection{LQR with Imperfect Model and $\hat L$}
 We now consider the deterministic model:
\begin{align}
\hat s_+ = \hat A s+\hat Ba. \label{eq:LQRModel}
\end{align}
In order to verify Theorem~\ref{thm:ZeTheorem}, we consider the stage cost
\begin{align*}
\hat L\left(s,a\right) &= Q_\star\left(s,a\right) - \gamma V_\star\left(\hat s_+\right) \\
&=  \tightMatr{c}{s \\ a}^\top\tightMatr{cc}{\hat T & \hat N \\ \star & \hat R}\tightMatr{c}{s\\ a}+ \left(1-\gamma\right)V_0.
\end{align*} 
This implies that $\hat T,\, \hat N,\,\hat R$ must satisfiy:
\begin{subequations}
\label{eq:hatL:Detail}
\begin{align}
\hat T + \gamma \hat A^\top S \hat A&= T+\gamma A^\top S A,  \\
\hat N + \gamma \hat A^\top S \hat B&= N+\gamma A^\top S B, \\
\hat R+\gamma \hat B^\top S \hat B &= R+\gamma B^\top S B.
\end{align}
\end{subequations}
The resulting value function 
reads as $\hat V\left(s\right) = s^\top \hat S s + \hat V_0$,
where $\hat S$ and the associated policy $\hat K$ satisfy:
\begin{subequations}
\label{eq:hatShatK}
\begin{align}
 \hat T + \gamma \hat A^\top \hat S \hat A &= \hat S + \left( \hat N+ \gamma \hat A^\top \hat S \hat B \right)\hat K,  \\
\left( \hat R + \gamma \hat B^\top \hat S \hat B \right)\hat K &= \hat N^\top+  \gamma \hat B^\top \hat S \hat A.
\end{align}
\end{subequations}
Using \eqref{eq:hatL:Detail}, we obtain that matrices $\hat S = S$ and $\hat K = K_\star$ satisfy \eqref{eq:hatShatK}. 

\subsection{Assumption~\eqref{eq:StabilityAssumption} in the LQR case}
In order for the Discrete Algebraic Riccati Equation (DARE)~\eqref{eq:hatShatK} to deliver a valid LQR solution $\hat S$, $\hat K$ (in the sense of a Bellman optimality backup), Assumption \eqref{eq:StabilityAssumption} needs to be satisfied. In the case $\gamma=1$, this requires that $\hat A-\hat B K_\star$ has all its eigenvalues inside the unit circle.
We illustrate this fact by the following simple example: 
\begin{align*}
	A=1, && B=1,&&T=1, &&R=2, && N=0,
\end{align*}
which yields $S=2$ and $K_\star=0.5$.
By using the model $\hat A = 2$, $\hat B = 1$,
we obtain
\begin{align*}
	\hat T=-5, && \hat R=2, && \hat N=-2,
\end{align*}
such that $\hat A-\hat B K_\star=1.5$ and $K_\star$ corresponds to the non-stabilizing solution $\hat S=2$ of the DARE. Note, however, that the DARE does have a stabilizing solution, which reads
\begin{align*}
	\hat S = 7, &&\hat K = 4/3 && \text{such that} &&\hat A-\hat B \hat K=2/3.
\end{align*}

%
%

\subsection{Economic LQR and Cost Modification $\varLambda(s,a)$}
We now consider an economic LQR, for which $\hat L(s,a)$ is indefinite. 
We introduce matrices $\delta T$, $\delta N$, $\delta R$ to define 
\begin{align*}
\varLambda(s,a) := \tightMatr{c}{s \\ a}^\top\tightMatr{ll}{\delta T & \delta N \\ \delta N^\top & \delta R}\tightMatr{c}{s \\ a},
\end{align*}
and observe that $\delta N = K_\star^\top \delta R$, $\delta R\succeq 0$ must hold in order for \eqref{eq:PiPreservation:Condition:L2} to be fulfilled, where $K_\star$ is the LQR controller gain associated to the stage cost $L$ for the true system~\eqref{eq:lqr_optimal_feedback}. By defining $\delta S := \delta T - K_\star^\top\delta R K_\star^{\phantom{\top}}$, we obtain
\begin{align*}
\E{}{\varLambda\left(s_+,\hat\pi(\hat s_+)\right)| s,a}= 
&\tightMatr{c}{s \\ a}^\top\tightMatr{cc}{\hat A^\top\delta S\hat A & \hat A^\top\delta S\hat B  \\ \hat B^\top\delta S\hat A  & \hat B^\top\delta S\hat B }\tightMatr{c}{s \\ a},
\end{align*}
such that admissible stage cost modifiers are based on the quadratic forms:
\begin{align*}
\delta W_L &=  \tightMatr{ll}{K_\star^\top\hspace{-1pt} \delta R K_\star & \hspace{-1pt}K_\star^\top \delta R \\ \delta R K_\star & \hspace{-1pt}\delta R}
- \tightMatr{ll}{ \gamma \hat A^\top\hspace{-1pt}\delta S\hat A -\delta S &  \hspace{-1pt}\gamma \hat A^\top\hspace{-1pt}\delta S\hat B  \\  \gamma \hat B^\top\hspace{-1pt}\delta S\hat A  &  \hspace{-1pt}\gamma \hat B^\top\hspace{-1pt}\delta S\hat B } \\
&=\delta W_L^0 + \delta W_L^1.
\end{align*}
It follows that the modified action-value function reads
\begin{align}
\hat Q(s,a) = \tightMatr{c}{s \\ a}^\top \left ( W + \delta W_L  \right ) \tightMatr{c}{s \\ a}.
\end{align}
Since we are interested in stability-enforcing schemes, one needs to choose $\delta R$ and $\delta S$ such that $W + \delta W_L \succ 0$ holds. Since this expression is linear in $\delta R$ and $\delta S$, the problem amounts to solving an LMI.

In the following, we state explicitly how the two contributions in $\delta W_L$ relate to conditions~\eqref{eq:PiPreservation:Condition:L2} and~\eqref{eq:strict_diss} respectively.
The term $\delta W_L^1$ can be framed as a quadratic stage cost rotation and the condition
\begin{align*}
	\delta W_L^1 = -\tightMatr{ll}{ \gamma \hat A^\top\delta S\hat A -\delta S &  \gamma \hat A^\top\delta S\hat B  \\  \gamma \hat B^\top\delta S\hat A  &  \gamma \hat B^\top\delta S\hat B } \succ 0,
\end{align*}
is the strict dissipativity condition~\eqref{eq:strict_diss} for the linear-quadratic case~\cite{Zanon2016b}. 
For $\delta R=0$, we obtain $\Lambda(s,a)=\lambda(s)=s^\top \delta S s$.

The term $\delta W_L^0$  resembles the stage cost used in~\cite{Chisci2001,Zanon2018} and satisfies
\begin{align*}
	\varLambda^0(s,a)=\tightMatr{c}{s \\ a}^\top \delta W_L^0 \tightMatr{c}{s \\ a} = (a+K_\star s)^\top\delta R(a+K_\star s),
\end{align*}
such that $\varLambda^0(s,a)>\varLambda^0(s,-K_\star s)$, $\forall \, a\neq -K_\star s$, $\forall \, s$. As discussed in~\cite{Chisci2001,Zanon2018}, the use of $\varLambda^0(s,a)$ as stage cost with a zero terminal cost yields a scheme which delivers the optimal feedback for the nominal model. The related value function and action-value function, however, are zero.

\section{Numerical Examples} \label{sec:Examples}
In this section, we propose two examples in order to illustrate the theoretical developments.

\subsection{Linear MPC} \label{sec:LinearMPC}
We first consider a simple linear MPC example to illustrate the methods above. We consider the MPC scheme:
\begin{subequations}
\label{eq:MPC:Example1}
\begin{align}
\min_{x,u}&\,\, V_0 + \frac{\gamma^N}{2}x_N^\top S_N x_N + \sum_{k=0}^{N-1}f^\top \matr{c}{x_k \\ u_k} \\
&\sum_{k=0}^{N-1} \frac{1}{2}\gamma^k\left(\left\|x_k\right\|^2 + \frac{1}{2}\left\|u_k\right\|^2 + w^\top s_k\right) \label{eq:MPC:Example1:Cost}\\
\mathrm{s.t.} &\quad x_{k+1} = Ax_k +  Bu_k + b, \label{eq:MPC:Example1:Model}\\
&\quad \matr{c}{\phantom{-}0 \\ -1} + \underline{x} -s_k\leq x_k \leq  \matr{c}{1 \\ 1} + \bar{x}+ s_k ,\label{eq:MPC:Example1:Constraints} \\
&\quad -1\leq u_k \leq 1, \qquad  x_0 = s,
\end{align}
\end{subequations}
where the NMPC parameters subject to the RL scheme are
\begin{align}
\theta = \left(V_0,\,\underline{x},\,\bar{x},\,b,\,f,\, A,\,B\right),
\end{align}
and where the $\ell_1$ relaxation uses the weight $w^\top = \matr{cc}{10^2 & 10^2}$. We selected a discount factor $\gamma = 0.9$. The terminal cost matrix $S_N$ is selected as the Riccati matrix underlying the LQR controller locally equivalent to the MPC scheme. The objective of the MPC scheme is to drive the states to the origin, such that the MPC reference is activating the lower bound of the first state. A horizon of $N=10$ is used. The MPC model is initially chosen as
\begin{align}
A = \matr{cc}{1 & 0.25\\ 0 & 1 },\quad B= \matr{c}{0.0312 \\ 0.25},
\end{align}
while all other parameters are initialized with zero values.

We will consider that the ``real" process is following the dynamics: 
\begin{align}
x_{k+1} = \matr{cc}{0.9 & 0.35\\ 0 & 1.1 }x_k +  \matr{c}{0.0813 \\ 0.2}u_k + \matr{c}{e_k \\ 0}
\end{align}
where $e_k$ is a random, uncorrelated, uniformly distributed variable in the interval $[-10^{-1},\, 0]$, and therefore drives the first state to violate its lower bound. 



\begin{figure}[t]
\begin{center}
	\includegraphics[width=0.45\textwidth,clip,trim=0 20 40 0]{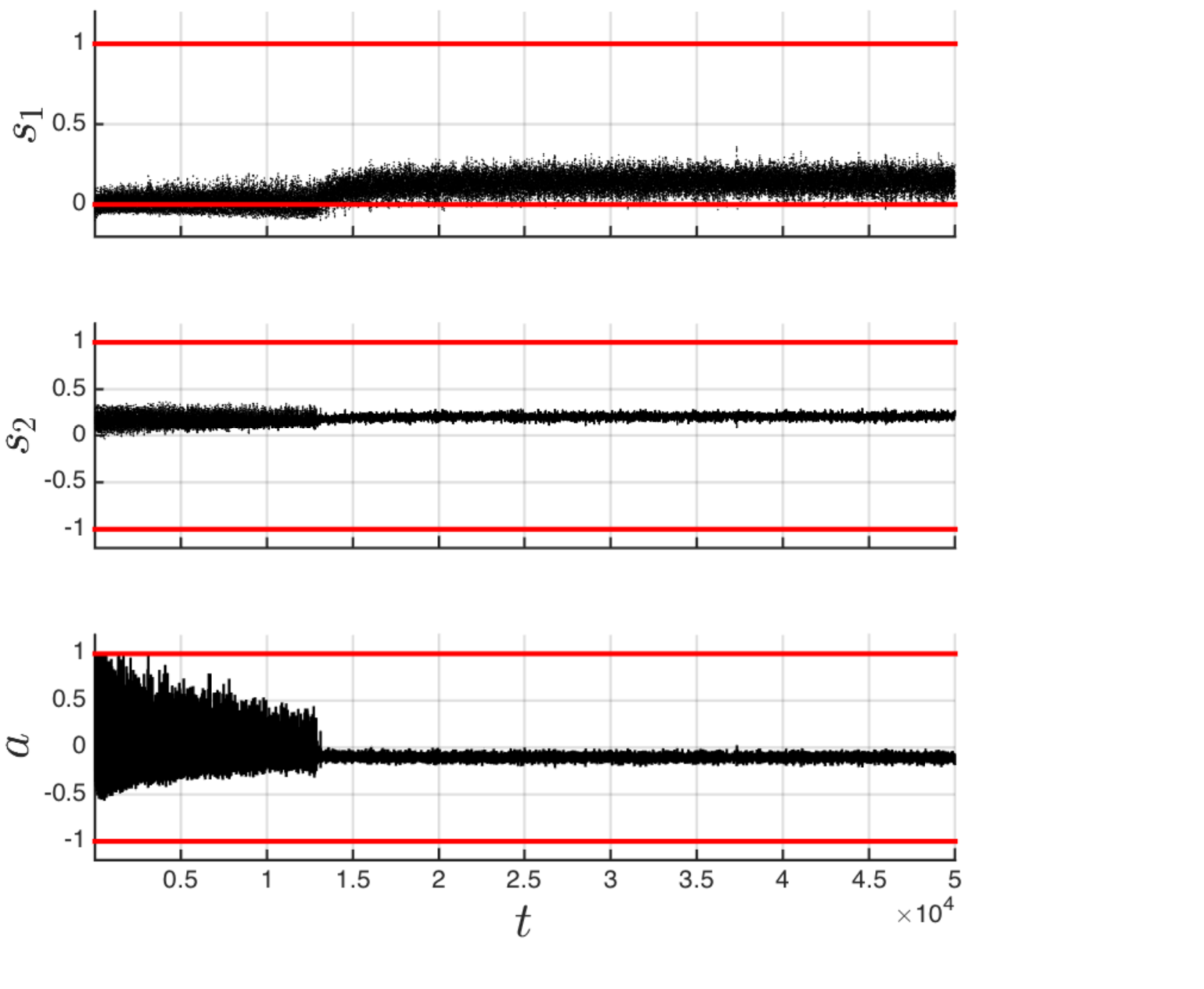}
\end{center}
\caption{State and input trajectories for the example detailed in Sec. \ref{sec:LinearMPC}.}
\label{fig:Altogether_Traj}
\end{figure}

\begin{figure}[t]
	\begin{center}
		\includegraphics[width=0.9\linewidth,clip,trim=0 20 90 0]{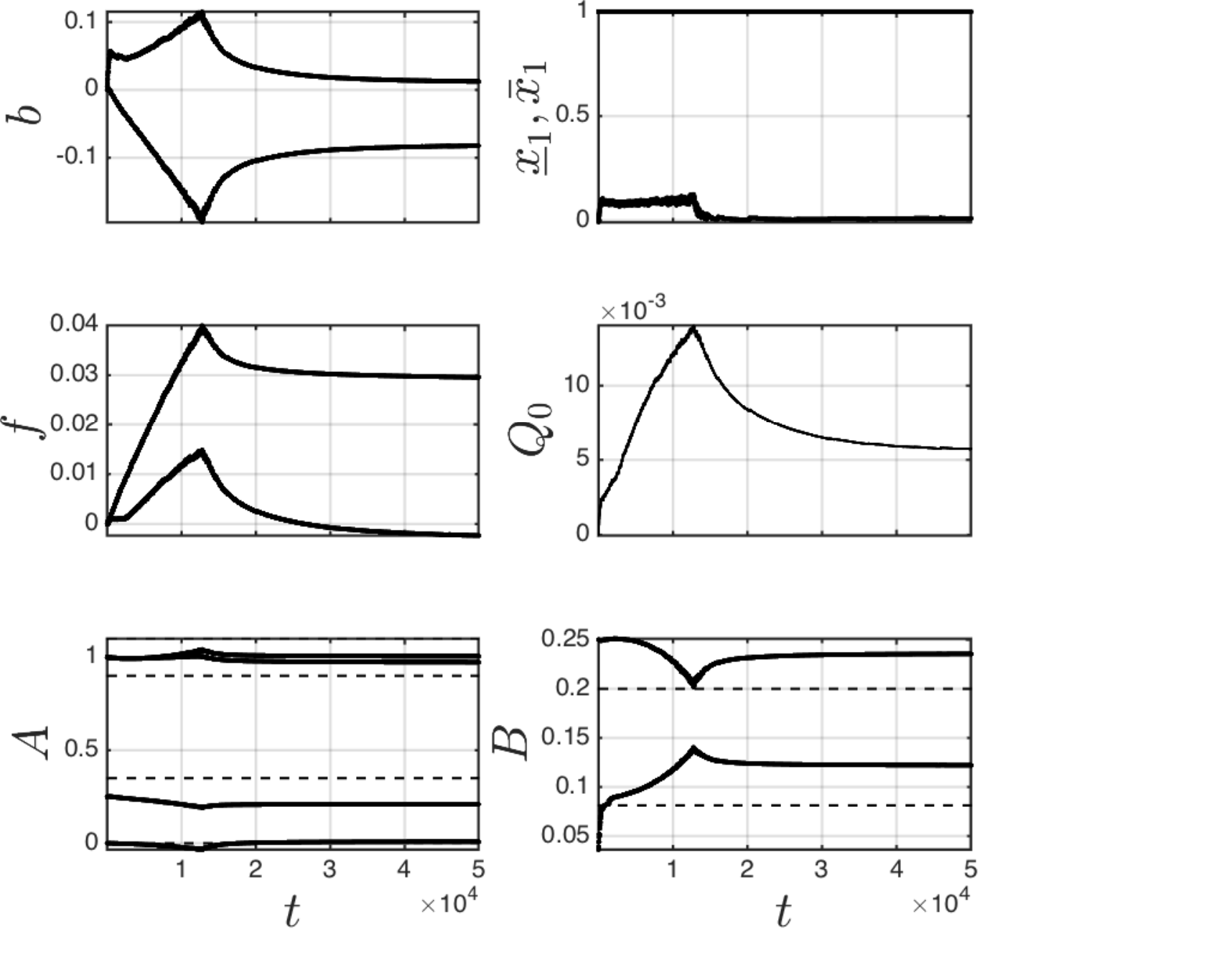}
	\end{center}
\caption{Trajectory of the parameters $\theta$ for the example detailed in Sec. \ref{sec:LinearMPC}. The dashed lines report the ``real system" $A$ and $B$ entries. One can observe that the RL scheme does not perform system identification, as the MPC model does not converge to the ``real system".}
\label{fig:Altogether_Param}
\end{figure}

\begin{figure}[t]
\begin{center}
	\includegraphics[width=0.45\textwidth,clip,trim=0 0 0 0]{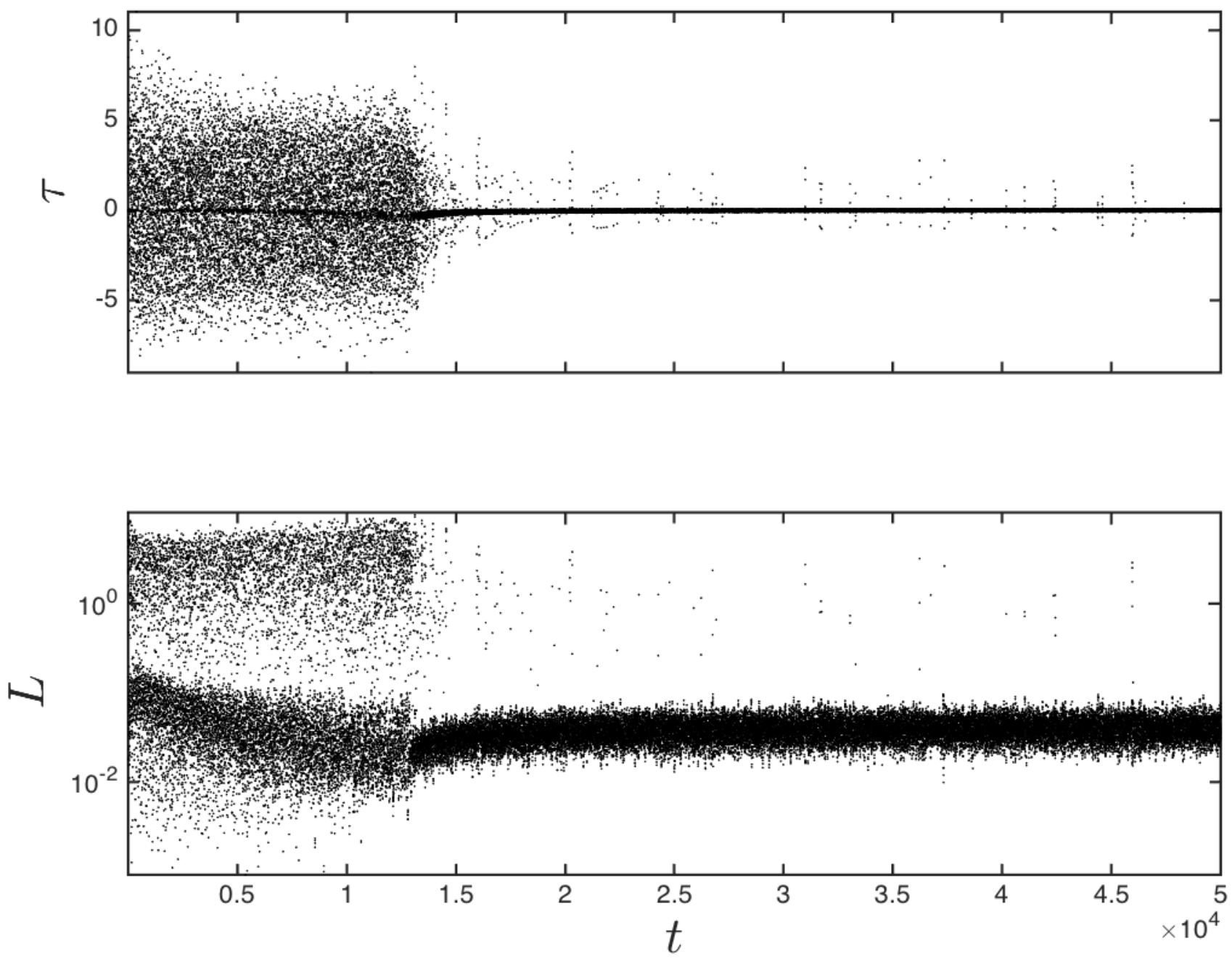}
\end{center}
\caption{Evolution of the stage cost \eqref{eq:MPC:Example1:Cost} and TD error \eqref{eq:TDError:OnPolicy} achieved for the example detailed in Sec. \ref{sec:LinearMPC}.}
\label{fig:Altogether_CostDelta}
\end{figure}

We use the on-policy algorithm \eqref{eq:Qlearning:OnPolicy} without introducing exploration. The step size was selected as $\alpha = 10^{-6}$. Figure~\ref{fig:Altogether_Traj} displays the resulting state and input trajectories. Figure~\ref{fig:Altogether_Param} shows the adaptation of the MPC parameters via the on-policy algorithm \eqref{eq:Qlearning:OnPolicy}. Figure~\ref{fig:Altogether_CostDelta} shows the evolution of the stage cost $L(s,a)$ (including the large $\ell_1$ penalties for the constraints violations), and the evolution of the TD error \eqref{eq:TDError:OnPolicy}. It can be observed that the RL algorithm manages to reduce the TD error $\tau$ to small values, averaging to zero. The state trajectories often violate the state bound $x_1\geq 0$ in the beginning, resulting in large control actions (see Figure~\ref{fig:Altogether_Traj}), but the RL adjusts the MPC parameters in order to avoid these expensive violations. The adaptation of the parameters is a combination of modifying the stage gradient $f$ and of introducing a model bias $b$ (mostly on the first state, subject to the process noise). 

We have also deployed this example without letting the RL scheme adapt the model parameters. The MPC performance is then increased mostly via tightening the bounds, and does not reach the performance displayed in Fig. \ref{fig:Altogether_Param}. For the sake of brevity, we do not report these results here.


\subsection{Evaporation Process}

We consider an example from the process industry, i.e. the evaporation process modeled in~\cite{Wang1994,Sonntag2006} and used in~\cite{Amrit2013a,Zanon2016b} in the context of economic MPC. 
The model equations include states $(X_2,P_2)$ (concentration and pressure) and controls $(P_{100},F_{200})$ (pressure and flow). The model further depends on concentration $X_1$, flow $F_2$ and temperatures $T_1,T_{200}$, which are assumed to be constant in the control model. In reality, these quantities are stochastic with variance $\sigma_{X_1}=1$, $\sigma_{F_1}=2$, $\sigma_{T_1}=8$, $\sigma_{T_{200}}=5$,  and mean centered on the nominal value. Bounds $(25,40)\leq (X_2,P_2) \leq (100,80)$ on the states and $u_\mathrm{l}=(100,100) \leq (P_{100},F_{200}) \leq (400,400)=u_\mathrm{u}$ on the controls are present. In particular, the bound $X_2\geq 25$ is introduced in order to ensure sufficient quality in the product. All state bounds are relaxed and translated into cost terms as in Section~\ref{sec:Qlearning}. The system dynamics are given by~\cite{Amrit2013a}
\begin{align}
M \dot X_2 &= F_1 X_1 - F_2 X_2, &
C \dot P_2 &= F_4-F_5,
\end{align}
where 
\begin{align*}
T_2 &= aP_2 + bX_2 + c, & T_3 &= dP_2 + e, \\
 \lambda F_4 &= Q_{100} - F_1C_\mathrm{p}(T_2-T_1), & T_{100} &= fP_{100} + g, \\
  Q_{100} &= U_{A_1}(T_{100}-T_2), & U_{A_1} &= h(F_1+F_3), \\
Q_{200} &= \frac{U_{A_2}(T_3-T_{200})}{1+U_{A_2}/(2C_\mathrm{p}F_{200})},& F_{100} &=\frac{Q_{100}}{\lambda_\mathrm{s}} , \\
\lambda F_5 &= Q_{200}, & F_2 &= F_1-F_4,
\end{align*}
and the model parameters are given in Table~\ref{tab:parameters}. 
The economic objective is given by
\begin{align*}
	L(x,u) = 10.09(F_2+F_3) + 600 F_{100} + 0.6 F_{200}.
\end{align*}

We introduce functions $\lambda_\theta, V^\mathrm{f}_\theta, l_\theta$ as fully parametrized quadratic functions defined by Hessian $H_\dagger$, gradient $h_\dagger$ and constant $c_\dagger$, $\dagger=\{\lambda, V^\mathrm{f}, l\}$ to formulate the ENMPC controller 
\begin{align*}
	\min_{z}\ \ & \lambda_\theta(x_0) + \gamma^N \left(V^\mathrm{f}_\theta(x_N) + w^\top \sigma_\mathrm{f}\right)\hspace{-14em} \\
	&\hspace{8em}+ \sum_{k=0}^{N-1} \gamma^k \left (l_\theta(x_k,u_k) + w^\top \sigma_k \right ) \hspace{-8em}\\
	\mathrm{s.t.} \ \ & x_{k+1} = f_\theta\left(x_k,u_k\right),&& x_0 = s, \\
	& g\left(u_k\right) \leq 0, && h_\theta\left(x_k\right) \leq \sigma_k.
\end{align*}
The model is parametrized as the nominal model with the addition of a constant, i.e. $f_\theta(x,u) = f(x,u) + c_f$. The control constraints are fixed and the state constraints are parametrized as simple bounds, i.e. $h(x)=(x-x_\mathrm{l},x_\mathrm{u}-x)$. The vector of parameter therefore reads as:
\begin{align*}
	\theta = ( H_\lambda, h_\lambda, c_\lambda, H_{V^\mathrm{f}}, h_{V^\mathrm{f}}, c_{V^\mathrm{f}}, H_l, h_l, c_l, c_f, x_\mathrm{l},x_\mathrm{u}).
\end{align*}
Constants $w=1$ are fixed and assumed to reflect the known cost of violating the state constraints.

\newcolumntype{Y}{>{\setlength\hsize{3,1em}\centering\arraybackslash}X}
\begin{table}
	\begin{tabularx}{0.5\textwidth}{c *{7}{Y}}
		\toprule
		$a$ & $b$ & $c$ & $d$ & $e$ & $f$ & $g$  \\
		0.5616 & 0.3126 & 48.43 & 0.507 & 55 & 0.1538 & 55 \\
		\midrule
		$h$ &$M$ & $C$ & $U_{A_2}$ & $C_\mathrm{p}$ & $\lambda$ & $\lambda_\mathrm{s}$ \\
		0.16 &20 & 4 & 6.84 & 0.07 & 38.5 & 36.6 \\
		\midrule
		$F_1$ & $X_1$ & $F_3$ & $T_1$ & $T_{200}$ &  &   \\
		10 & 5 \% & 50 &       40 &        25      &  &  \\
		\bottomrule
	\end{tabularx}
	\caption{Model Parameters. The units are omitted and are consistent with the physical quantities they correspond to.}
	\label{tab:parameters}
\end{table}

\begin{figure}
	\includegraphics[width=0.95\linewidth,clip,trim= 0 150 0 120]{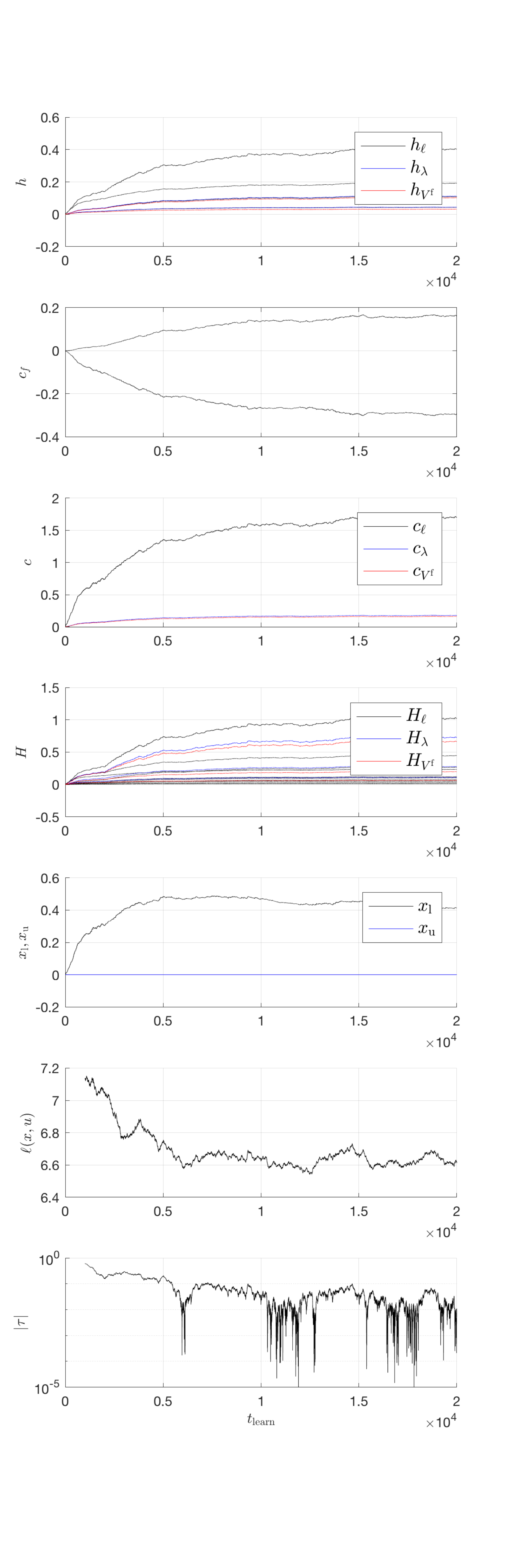} 
	\caption{Evolution of the parameters (increment w.r.t. the initial guess value) and of the TD-error (averaged over the preceding $1000$ samples) during learning.}
	\label{fig:nmpc_learn}
\end{figure}

We use the batch policy update~\eqref{eq:Qlearning:OnPolicy} with $\alpha=10^{-4}$ and we update the parameters with the learned ones every $N_\mathrm{upd}=2000$ time steps. In order to induce enough exploration, we use an $\epsilon$-greedy policy which is greedy $90\ \%$ of the samples, while in the remaining $10 \ \%$ we perturb the optimal feedback as
\begin{align*}
	a = \mathrm{sat}(u_0^* + e,u_\mathrm{l},u_\mathrm{u}), && e \sim \mathcal{N}(0,1),
\end{align*}
where $\mathrm{sat}(\cdot,u_\mathrm{l},u_\mathrm{u})$ saturates the input between its lower and upper bounds $u_\mathrm{l}=(100,100),u_\mathrm{u}=(400,400)$ respectively.
We initialize the ENMPC scheme by tuning the cost using the economic-based approach proposed in~\cite{Zanon2016b}, which is based on the nominal model. For the model we use $c_f=0$ and the bounds are initialized at their nominal values.
While at every step we do check that $H_l, H_{V^\mathrm{f}}\succ 0$, during the learning phase the parameters never violate this constraint. As displayed in Figure~\ref{fig:nmpc_learn}, the algorithm converges to a constant parameter value while reducing the average TD-error.

\begin{figure}
	\includegraphics[width=0.9\linewidth]{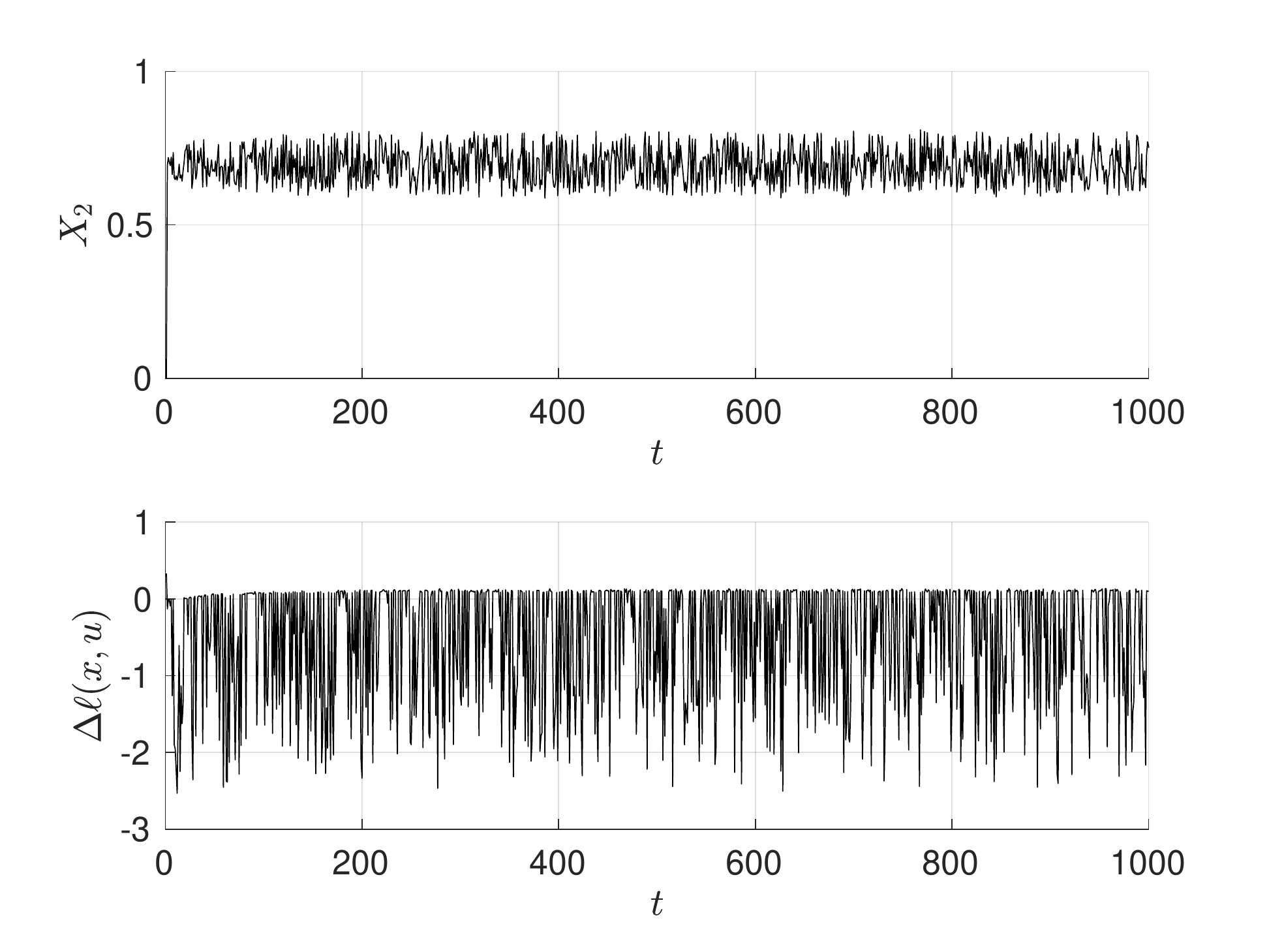}
	\caption{Closed-loop simulations: difference between the RL-tuned NMPC, and the  nominal economic tuning~\cite{Zanon2016b}. 
	Top plot: difference in concentration $X_2$.
	Bottom plot: difference in economic cost. 
	}
	\label{fig:nmpc_cl}
\end{figure}

We ran a closed-loop simulation to compare the performance of RL-tuned NMPC with the one 
using the economic-based tuning proposed in~\cite{Zanon2016b}. We display in Figure~\ref{fig:nmpc_cl} the difference in cost and concentration $X_2$ between the two ENMPC schemes. It can be seen that RL keeps $X_2$ at higher values in order to reduce the violation of the quality constraint. This entails an improvement in the cost which is about $7\%$ in the considered scenario.


It is important to stress here that both the cost and the model parametrization do not have the same structure as the real cost and model. Therefore, the action-value function and, consequently, the policy can be learned only approximately. Another important remark concerns the role of the storage function $\lambda_\theta$. RL exploits this function to keep the stage cost positive-definite. If we invert the cost modification~\eqref{eq:PiPreservation} in order to recover an approximation of the economic stage cost, we obtain an indefinite one.

We ran an additional simulation in which the model used in simulations is deterministic and coincides with the one used for predictions in NMPC. We initialize the learning phase with the nominally tuned parameters from~\cite{Zanon2016b}. In this case, RL keeps the parameters essentially unaltered, which suggests that the nominal tuning was already optimal.

Consider the naive initial guess $H_l=I$, $x_\mathrm{l}=(25,40)$, $x_\mathrm{u}=(100,80)$, while all other parameters are $0$. In order to obtain parameters which are close to convergence in each batch, we set $N_\mathrm{upd}=20000$. In this case, with $10^5$ samples we did not yet obtain convergence. Even though convergence of the RL scheme is much slower, the parameters are approaching a steady value and the closed-loop performance is improved with respect to the initial guess.


While our simulation results are promising, we ought to specify here that we have observed some potential difficulties related - to the best of our knowledge - to (i) using a gradient method in adjusting the ENMPC parameters and (ii) using a $Q$-learning method as a proxy for learning the optimal policy. These observations arguably call for exploring the effectiveness of deploying $2^\mathrm{nd}$-order methods, such as e.g. LSTD-type methods \cite{Sutton1998}, and policy gradient approaches \cite{Silver2014} to alleviate the difficulties observed in some cases.

\section{Conclusions}
In this paper, we propose to use Economic NMPC schemes to support the parametrization of the value functions and/or the policy which is an essential component of Reinforcement Learning. We show that the Economic NMPC schemes can generate the optimal policy for the real system even if the underlying model is wrong, by adjusting the stage and terminal cost alone. We also show how a positive stage and terminal cost can be used in the ENMPC scheme to learn the optimal control policy, resulting in an ENMPC scheme that is stable by construction. The resulting ENMPC delivers the optimal control policy for the real system if that policy is itself stabilizing. We additionally detail how some classic RL methods can be deployed in practice to adjust the parameters of the ENMPC scheme. The methods are illustrated in simulations.

Future work will propose improvements in the Reinforcement Learning algorithms specific for ENMPC, and propose an efficient combination of the existing classic model-tuning techniques and the Reinforcement-Learning-based tuning. 

\bibliographystyle{plain}
\bibliography{syscop}

\end{document}